\documentclass[12pt]{article}
\usepackage{amsmath}

\usepackage[letterpaper, portrait, margin=1in]{geometry}

\begin{document}

\pagestyle{plain}

\topmargin 0in
\oddsidemargin 0in
\evensidemargin 0in
\headheight 0in
\headsep 0in
\topskip 0in

\title{Isothermal and Adiabatic Elastic Tensors}
\author{Michael J. Waters and Andrew W. Bielawski}
\maketitle

\begin{abstract}
The adiabatic elastic modulus is often useful in the high frequency response of materials. Unfortunately, it can be much more difficult to directly measure the adiabatic elastic modulus of material than the isothermal elastic modulus. We derive the relationship between the adiabatic and isothermal elastic tensors from the first law of thermodynamics.
\end{abstract}

\section{Notation}

\begin{align*}
\sigma_{kl}&=\text{Stress Tensor}
\\
\varepsilon^o_{ij}&=\text{Strain Tensor, includes both stress-induced and temperature-induced strain}
\\
T&=\text{Temperature}
\\
S&=\text{Entropy}
\\
U&=\text{Internal Energy}
\\
H&=\text{Enthalpy}
\\
p&=\text{Pressure}
\\
V&=\text{Volume}
\\
\alpha_{ij}&=\text{Coefficient of Thermal Expansion}
\\
C_{ijkl}&=\text{Stiffness Tensor}
\\
S_{ijkl}&=\text{Compliance Tensor}
\\
c_{\sigma_{ij}}&=\text{Heat Capacity at Constant Stress}
\\
c_{\varepsilon^o_{ij}}&=\text{Heat Capacity at Constant Strain}
\end{align*}

\section{Definitions}

\subsection{Heat Capacity}

With liquids, there are two possible heat capacities which can be defined, one at constant pressure and one at constant volume. We are defining an analogous pair of heat capacities at constant stress and at constant strain. The constant strain heat capacity is defined as:

\begin{equation} \label{heat_capacity_constant_strain} 
\left(\frac{\partial U}{\partial T}  \right)_{ \varepsilon^o_{ij}} \equiv c_{\varepsilon^o_{ij}}
\end{equation}
\\
The constant stress heat capacity is defined as:

\begin{equation} \label{heat_capacity_constant_stress}
\left(\frac{\partial H}{\partial T}  \right)_{ \sigma_{ij}} \equiv c_{\sigma_{ij}}
\end{equation}

\subsection{Thermal Expansion}
The differential strain of a material is defined by changes in temperature and stress state. 

\begin{equation} 
d \varepsilon^o_{ij} \equiv \left(\frac{\partial \varepsilon^o_{ij}}{\partial T}  \right)_{\sigma_{kl}} d T+ \left( \frac{\partial \varepsilon^o_{ij}  }{\partial \sigma_{kl}} \right)_T d\sigma_{kl}
\end{equation}
The thermal expansion coefficient is defined at constant stress state.

\begin{equation} \label{def_thermal_expansion} 
\left(\frac{\partial \varepsilon^o_{ij}}{\partial T}  \right)_{\sigma_{kl}}   \equiv   \big (\alpha_{ij}  \big )_{\sigma_{kl}}
\end{equation}

\subsection{Isothermal Stiffness and Compliance}
The isothermal stiffness tensor is defined as:
\begin{equation} 
 \left( \frac{\partial \sigma_{ij}  }{\partial  \varepsilon^o_{kl} } \right)_T \equiv C_{ijkl}
\end{equation}
Correspondingly the isothermal compliance is defined as:

\begin{equation}
 \left( \frac{\partial \varepsilon^o_{ij}  }{\partial \sigma_{kl}} \right)_T
 \equiv S_{ijkl}
\end{equation}

\subsection{Differential Form of Internal Energy and Free Energies}
Our differential definition of internal energy with stress and strain work is:
\begin{equation} \label{dU}
dU=TdS+\sigma_{ij}d\varepsilon^o_{ij}
\end{equation}
The internal energy relations to our enthalpy, Helmholtz free energy, and Gibbs free energy are:

\begin{align}
H & = U - \sigma_{ij}\varepsilon^o_{ij} \\
A & = U - TS  \\
G & = U - TS -  \sigma_{ij}\varepsilon^o_{ij}
\end{align}
 In differential form:
\begin{align}
dH & = TdS - \varepsilon^o_{ij} d \sigma_{ij}  \label{dH} \\
dA & = - SdT + \sigma_{ij}d\varepsilon^o_{ij}   \label{dA} \\
dG & = - SdT - \varepsilon^o_{ij} d \sigma_{ij}  \label{dG}
\end{align}

\section{Introduction}
Pressure waves such as sound waves and seismic waves are generally well understood phenomena but over many years and scientific fields have increasingly become useful as a means of probing the structure and behavior in a wide range system sizes from Brillouin light scattering techniques to asteroseismology. The basic physical model for any wave is the wave equation. Pressure waves are mechanical and their wave equations can be derived purely by applying Lagrange's second equation to the thermodynamic equation of state and a continuity equation. The results of which for isotropic media look something like this:
\begin{equation}
\frac{\partial^{2} p}{\partial t^{2}}=\left ( \frac{\partial p}{\partial \rho} \right )_{S} \nabla^{2}p, \notag
\end{equation}
where $\left ( \frac{\partial p}{\partial \rho} \right )_{S} = c^2$ the propogation speed.  In an ideal gas, it is trivial to show that this only depends on the temperature. 
in an anisotropic material using Einstein notation, they look like this:
\begin{equation}
 \partial_{tt}u_i = \frac{1}{\rho}
 \left( \frac{\partial \sigma_{ij}  }{\partial  \varepsilon^o_{kl} } \right)_S
\partial_j \partial_l u_k
\notag
\end{equation}
where $u_i$ is the displacement. The adiabatic stiffness tensor in the parenthesis can be dificult to measure so we seek in this derivation to related it to the more readily available isothermal stiffness tensor.

\subsection{Derivation}

We start a triple product rule that relates the adiabatic stiffness tensor to two other quantities:

\begin{equation} \label{eq:1} 
\left(\frac{\partial \sigma_{kl}}{\partial \varepsilon^o_{ij}}  \right)_S
\left(\frac{\partial \varepsilon^o_{ij}}{\partial S} \right)_{\sigma_{kl}}
\left(\frac{\partial S}{\partial \sigma_{kl}}  \right)_{\varepsilon^o_{ij}}
= -1
\end{equation}

\begin{large}
$ \left(\frac{\partial \sigma_{kl}}{\partial \varepsilon^o_{ij}}  \right)_S $
\end{large}
is the adiabatic stiffness tensor. The goal of this derivation is to find an expression that relates this term to the isothermal stiffness tensor. To this end, we must find expressions for the other terms in Equation~\ref{eq:1}.

We first find an expression for the second term in Equation~\ref{eq:1},
\begin{large}
$\left(\frac{\partial \varepsilon^o_{ij}}{\partial S} \right)_{\sigma_{kl}}$.
\end{large}
This term represents the change in strain due to a change in entropy at constant stress. We can expand this term using the chain rule:

\begin{equation} \label{eq:2} 
\left(\frac{\partial \varepsilon^o_{ij}}{\partial S} \right)_{\sigma_{kl}}=
\left(\frac{\partial \varepsilon^o_{ij}}{\partial T}  \right)_{\sigma_{kl}}
\left(\frac{\partial T}{\partial S}  \right)_{\sigma_{kl}}
\end{equation}

The first term on the right hand side of Equation~\ref{eq:2} is simply the coefficient of thermal expansion at constant stress as defined in Equation \ref{def_thermal_expansion}:
\begin{equation}
\left(\frac{\partial \varepsilon^o_{ij}}{\partial T}  \right)_{\sigma_{kl}}   \equiv   \big (\alpha_{ij}  \big )_{\sigma_{kl}} \notag
\end{equation}

We must find an expression for the second term on the right hand side of Equation~\ref{eq:2},
\begin{large}
$\left(\frac{\partial T}{\partial S}  \right)_{\sigma_{kl}}$.
\end{large}
This is the change in temperature due to change in entropy at constant stress. In order to obtain this expession, we use the chain rule to expand the change in enthalpy due to a change in entropy at constant stress:

\begin{equation} \label{eq:4} 
\left( \frac{\partial H}{\partial S} \right) _{\sigma_{kl}}=
\left( \frac{\partial H}{\partial T}  \right) _{\sigma_{kl}}
\left (\frac{\partial T}{\partial S}  \right) _{\sigma_{kl}}
\end{equation}

In order to solve Equation~\ref{eq:4} for
\begin{large}
$\left(\frac{\partial T}{\partial S}  \right) _{\sigma_{kl}}$,
\end{large}
we must find expressions for the first two terms in Equation~\ref{eq:4}. We start with an expression for enthalpy in differential form:

\begin{equation} \label{eq:5} 
dH = TdS - \varepsilon^o_{kl}d\sigma_{kl}
\end{equation}

Since we are under the condition of constant stress, Equation~\ref{eq:5} can be solved assuming $d\sigma_{kl}=0$:

\begin{equation} \label{eq:6} 
\left(\frac{\partial H}{\partial S} \right)_{\sigma_{kl}}=T
\end{equation}

The first term on the right hand side of Equation~\ref{eq:4} is the heat capacity at constant stress as defined in Equation \ref{heat_capacity_constant_stress}.

\begin{equation} \label{eq:7} 
\left(\frac{\partial H}{\partial T}  \right)_{\sigma_{kl}}\equiv c_{\sigma_{kl}} \notag
\end{equation}

Substituting Equations \ref{heat_capacity_constant_stress} and \ref{eq:6} back into Equation~\ref{eq:4}:

\begin{align} \label{eq:8} 
T &= c_{\sigma_{kl}}\left(\frac{\partial T}{\partial S}  \right)_{\sigma_{kl}}\notag\\
\left(\frac{\partial T}{\partial S}  \right)_{\sigma_{kl}} &= \frac{T}{c_{\sigma_{kl}}}
\end{align}

We can now substitute Equations~\ref{def_thermal_expansion} and~\ref{eq:8} into Equation~\ref{eq:2} to obtain the second term in Equation~\ref{eq:1}:

\begin{equation} \label{eq:9} 
\left(\frac{\partial \varepsilon^o_{ij}}{\partial S} \right)_{\sigma_{kl}}=
\big (\alpha_{ij}  \big )_{\sigma_{kl}}
\left(\frac{T}{c_{\sigma_{kl}}} \right)
\end{equation}

Next we must find an expression for the third term in Equation~\ref{eq:1},\begin{large}
$\left(\frac{\partial S}{\partial \sigma_{kl}}  \right)_{\varepsilon^o_{ij}}$,
\end{large}
which represents the change in entropy due to a change in stress at constant strain. An alternative expression for this term can be found using the Maxwell relation from internal energy:

\begin{equation} \label{eq:10} 
\left(\frac{\partial T}{\partial \varepsilon^o_{ij}} \right)_S=
\left(\frac{\partial \sigma_{ij}}{\partial S} \right)_{\varepsilon^o_{ij}}
\end{equation}

We use again use the triple product rule, this time including the change in temperature due to a change in strain at constant entropy:

\begin{equation} \label{eq:11} 
\left(\frac{\partial T}{\partial \varepsilon^o_{ij}} \right)_S
\left(\frac{\partial S}{\partial T} \right)_{\varepsilon^o_{ij}}
\left(\frac{\partial \varepsilon^o_{ij}}{\partial S} \right)_T=-1
\end{equation}

Next we find an expression for the second term in Equation~\ref{eq:11},
\begin{large}
$\left(\frac{\partial S}{\partial T} \right)_{\varepsilon^o_{ij}}$,
\end{large}
which represents the change in entropy due to a change in temperature at constant strain. We start by using the chain rule to expand,
\begin{large}
$\left(\frac{\partial U}{\partial S} \right)_{\varepsilon^o_{ij}}$,
\end{large}
which is the change in internal energy due to a change in entropy at constant strain:

\begin{equation} \label{eq:12} 
\left(\frac{\partial U}{\partial S} \right)_{\varepsilon^o_{ij}}=
\left(\frac{\partial U}{\partial T} \right)_{\varepsilon^o_{ij}}
\left(\frac{\partial T}{\partial S} \right)_{\varepsilon^o_{ij}}
\end{equation}

We use an expression for internal energy in differential form:

\begin {equation} \label{eq:13} 
dU = TdS - pdV
\end{equation}

Since strain is held constant, we assume that volume is also constant ($dV=0$):

\begin{align} \label{eq:14} 
dU &= TdS \notag\\
\left(\frac{\partial U}{\partial S} \right)_{\varepsilon^o_{ij}} &= T
\end{align}

The first term on the right hand side of Equation~\ref{eq:12},
\begin{large}
$\left(\frac{\partial U}{\partial T} \right)_{\varepsilon^o_{ij}}$,
\end{large}
was defined in Equation \ref{heat_capacity_constant_strain} as the heat capacity at constant strain:

\begin{equation} \label{eq:15} 
\left(\frac{\partial U}{\partial T} \right)_{\varepsilon^o_{ij}} \equiv c_{\varepsilon^o_{ij}} \notag
\end{equation}

Substituting Equations \ref{heat_capacity_constant_strain} and \ref{eq:14} into Equation~\ref{eq:12} gives us an expression for the second term in Equation~\ref{eq:11}:

\begin{align} \label{eq:16} 
T &= c_{\varepsilon^o_{ij}}\left(\frac{\partial T}{\partial S} \right)_{\varepsilon^o_{ij}} \notag\\
\left(\frac{\partial T}{\partial S} \right)_{\varepsilon^o_{ij}} &= \frac {T}{c_{\varepsilon^o_{ij}}} \notag\\
\left(\frac{\partial S}{\partial T} \right)_{\varepsilon^o_{ij}} &= \frac {c_{\varepsilon^o_{ij}}}{T}
\end{align}

Next we find an expression for the third term in Equation~\ref{eq:11}. We start with the Maxwell relation for Helmholtz free energy:

\begin{equation} \label{eq:17} 
\left(\frac{\partial S}{\partial \varepsilon^o_{ij}} \right)_T=
-\left(\frac{\partial \sigma_{ij}}{\partial T}  \right)_{\varepsilon^o_{ij}}
\end{equation}

Strain can be written in differential form as:

\begin{equation} \label{eq:18} 
d\varepsilon^o_{ij} = S_{ijkl}d\sigma_{kl} + \alpha_{ij}dT
\end{equation}

Since the strain is constant ($d\varepsilon^o_{ij}=0$):

\begin{align} \label{eq:19}
0& = S_{ijkl}d\sigma_{kl} + \alpha_{ij}dT \notag \\
-S_{ijkl}d\sigma_{kl}& = \alpha_{ij}dT \notag \\
-S_{ijkl} \left(\frac {\partial \sigma_{kl}}{\partial T} \right)_{\varepsilon^o_{ij}}& = \alpha_{ij}
\end{align}

We multiply both sides of Equation~\ref{eq:19} by the stiffness tensor, $C_{klij}$:

\begin{align} \label{eq:20} 
(C_{klij})(-S_{ijkl}) \left(\frac {\partial \sigma_{kl}}{\partial T} \right)_{\varepsilon^o_{ij}}& = C_{klij} \alpha_{ij} \notag \\
-\left(\frac {\partial \sigma_{kl}}{\partial T} \right)_{\varepsilon^o_{ij}}& = C_{klij}\alpha_{ij}
\end{align}

Substituting Equation~\ref{eq:17} into Equation~\ref{eq:20} and swapping indices:

\begin{align} \label{eq:21} 
\left(\frac{\partial S}{\partial \varepsilon^o_{ij}} \right)_ T& = C_{ijkl}\alpha_{kl} \notag \\
\left(\frac{\partial \varepsilon^o_{ij}}{\partial S} \right)_T& =(C_{ijkl}\alpha_{ij})^{-1}
\end{align}

We now substitute Equations~\ref{eq:16} and~\ref{eq:21} into Equation~\ref{eq:11}:

\begin{align}
\left(\frac{\partial T}{\partial \varepsilon^o_{ij}} \right)_S \left(\frac{c_{\varepsilon^o_{ij}}}{T} \right) (C_{ijkl}\alpha_{ij})^{-1}& = -1 \notag \\
\left(\frac{\partial T}{\partial \varepsilon^o_{ij}} \right)_S& = -\left(\frac {T}{c_{\varepsilon^o_{ij}}} \right) (C_{ijkl}\alpha_{ij}) \notag
\end{align}

Again using the Maxwell relation from Equation~\ref{eq:10}:

\begin{align} \label{eq:22} 
\left(\frac{\partial \sigma_{ij}}{\partial S} \right)_{\varepsilon^o_{ij}}&=\left(\frac{\partial T}{\partial \varepsilon^o_{ij}} \right)_S=-\left(\frac {T}{c_{\varepsilon^o_{ij}}} \right)(C_{ijkl}\alpha_{ij}) \notag \\
\left(\frac{\partial S}{\partial \sigma_{ij}} \right)_{\varepsilon^o_{ij}}&=-\left(\frac {c_{\varepsilon^o_{ij}}}{T} \right)(C_{ijkl}\alpha_{ij})^{-1}
\end{align}

We can now substitute Equations~\ref{eq:9} and~\ref{eq:22} back into Equation~\ref{eq:1}:

\begin{gather} \label{eq:23} 
\left(\frac{\partial \sigma_{kl}}{\partial \varepsilon^o_{ij}}  \right)_S (\alpha_{ij})_{\sigma_{kl}}\left(\frac{T}{c_{\sigma_{kl}}} \right)\Bigg[-\left(\frac {c_{\varepsilon^o_{ij}}}{T} \right)(C_{ijkl}\alpha_{ij})^{-1}\Bigg]=-1 \notag \\
\left(\frac{\partial \sigma_{kl}}{\partial \varepsilon^o_{ij}}  \right)_S=\left(\frac {c_{\sigma_{kl}}}{c_{\varepsilon^o_{ij}}} \right)C_{ijkl}
\end{gather}

\section{References}
\section{Appendices}
\subsection{Maxwells Relations}
The Maxwell relation for the internal energy $U$ (Equation~\ref{dU}) is:
\begin{equation} \label{maxwell_U}
\left ( \frac{\partial T}{\partial \varepsilon^{o}_{ij}} \right )_{S,\varepsilon^{o}_{kl\neq ij}}=\left ( \frac{\partial \sigma_{ij}}{\partial S} \right )_{\varepsilon^{o}_{kl}}
\end{equation}
On the left, all strains are held constant except $\varepsilon^{o}_{ij}$.
\\
\\
Likewise, the Maxwell relation for enthalpy $H$ (Equation~\ref{dH}) is:
\begin{equation} \label{maxwell_H}
\left ( \frac{\partial T}{\partial \sigma_{ij}} \right )_{S,\sigma_{kl\neq ij}}=-\left ( \frac{\partial \varepsilon^{o}_{ij}}{\partial S} \right )_{\sigma_{kl}}
\end{equation}
\\
The Maxwell relation for Helmholtz free energy $A$ (Equation~\ref{dA}) is:
\begin{equation} \label{maxwell_A}
-\left ( \frac{\partial S}{\partial \varepsilon^{o}_{ij}} \right )_{T,\varepsilon^{o}_{kl\neq ij}}=\left ( \frac{\partial \sigma_{ij}}{\partial T} \right )_{\varepsilon^{o}_{kl}}
\end{equation}
\\
The Maxwell relation for Gibbs free energy $A$ (Equation~\ref{dG}) is:
\begin{equation} \label{maxwell_G}
\left ( \frac{\partial S}{\partial \sigma_{ij}} \right )_{T,\sigma_{kl\neq ij}}=\left ( \frac{\partial \varepsilon^{o}_{ij}}{\partial T} \right )_{\sigma_{kl}}
\end{equation}

\end{document}